# Mitigating Denial-of-Service Attacks Using Secure Service Overlay Model

Shalaka S. Chowriwar [#1], Madhulika S. Mool[#2], Prajyoti P.Sabale[#3], Sneha S.Parpelli[#4], Mr.Nilesh Sambhe [#5]

[#]*Department of Computer Technology, Yeshwantrao Chavan College Of Engineering , Nagpur ,India.*

## ABSTRACT

*Denial of service (DoS) and Distributed Denial of Service (DDoS) attacks continue to threaten the reliability of networking systems. Previous approaches for protecting networks from DoS attacks are reactive in that they wait for an attack to be launched before taking appropriate measures to protect the network. This leaves the door open for other attacks that use more sophisticated methods to mask their traffic. A secure overlay services (SOS) architecture has been proposed to provide reliable communication between clients and a target under DoS attacks. The SOS architecture employs a set of overlay nodes arranged in three hierarchical layers that controls access to the target. We propose an architecture called secure overlay services (SOS) that proactively prevents denial of service (DoS) attacks, which works toward supporting emergency services, or similar types of communication. The architecture uses a combination of secure overlay tunneling, routing via consistent hashing, and filtering. We reduce the probability of successful attacks by: 1) performing intensive filtering near protected network edges, pushing the attack point into the core of the network, where high-speed routers can handle the volume of attack traffic and 2) introducing randomness and anonymity into the forwarding architecture, making it difficult for an attacker to target nodes along the path to a specific SOS-protected destination. Using simple analytical models, we evaluate the likelihood that an attacker can successfully launch a DoS attack against an SOS protected network. Our analysis demonstrates that such an architecture reduces the likelihood of a successful attack to minuscule levels.*

## I.INTRODUCTION

A secure system meets or exceeds an application-specified set of security policy *requirements*. For example, in message delivery, the high-level requirements may be that the correct information gets to the right person, in the right place, *at the right time*. The details of "right" are determined by the application's needs. One threat to timely data delivery in a public network such as the Internet is denial of service (DoS) attacks: these attacks overwhelm the processing or link capacity of the target site (or routers that are topologically close) by saturating it (them) with bogus packets. Such attacks can disrupt legitimate communications at minimal cost and danger to the attacker, as has been demonstrated repeatedly in recent years.In the SOS architecture, we address the problem of securing communication in today's existing Internet protocol (IP) infrastructure from denial of service (DoS) attacks, where the communication is between a predetermined location and a set of well-known users, located anywhere in the wide-area network, who have authorization to communicate with that location. We focus our efforts on protecting a site that stores information that is difficult to replicate due to security concerns or due to its dynamic nature. An example is a database that maintains timely or confidential information such as building structure reports, intelligence, assignment updates, or strategic information. We assume that there is a predetermined set of clients scattered throughout the network who require access to this information, from anywhere in the network. The portion of the network immediately surrounding the target (location to be protected) aggressively filters and blocks all incoming packets whose source addresses are not "approved." The small set of source addresses (potentially as small as 2–3 addresses) that are "approved" at any particular time is kept secret so that attackers cannot use them to pass through the filter. These addresses are picked from among those within a distributed set of nodes throughout the wide area network, that form a secure overlay: any transmissions that wish to traverse the overlay must first be validated at entry points of the overlay[7]. Once inside the overlay, the traffic is tunneled securely for several hops along the overlay to the "approved" (and secret from attackers) locations, which can then forward the validated traffic through the filtering routers to the target. The two main principles behind our design are: 1) elimination of communication points, which constitute attractive DoS targets, via a combination of filtering and overlay routing to obscure the identities of the sites whose traffic is permitted to pass through the filter and 2) the ability to recover from random or induced failures within the forwarding infrastructure or within the secure overlay nodes. The attackers can also know the IP addresses of the nodes that participate in the overlay and of the target that is to be protected, as well as the details of the operation of protocols used to perform the forwarding. However, we assume that: 1) the attacker does not have unobstructed access to the network core and 2) the attacker cannot severely disrupt large parts of the backbone[3][4].

## II.DoS AND DDoS OVERVIEW

### A.Denial-of- Service (DoS) Attacks :

A DoS attack is a malicious attempt by a single person or a group of people to disrupt an online service. DoS attacks can be launched against both services, e.g., a web server, and networks, e.g., the network connection to a server. The impact of DoS attacks can vary from minor inconvenience to users of a website, to serious financial losses for companies that rely on their on-line availability to do business. As emergency and essential services become reliant on the Internet as part of their communication infrastructure, the consequences of DoS attacks could even become life-threatening. Hence, it is crucial to deter, or otherwise minimize, the damage caused by DoS attacks. Types of DoS attacks are  TCP SYN Flood Attack,  UDP Flood Attacks, Ping of Death Attacks,  Smurf Attacks,  Teardrop Attacks,  Bonk Attacks, Land Attacks[6].

### B.Distributed Denial of Service (DDoS) Attacks :

When an attacker attacks from multiple source systems, it is called a Distributed Denial of Service (DDoS) attack. If the attacker is able to organize a large amount of users to connect to the same website at the same time, the webserver, often configured to allow a maximum number of client connections, will deny further connections. Hence, a denial of service will occur.  However, the attacker typically does not own these computers. The actual owners are usually not aware of their system being used in a DDoS attack. The attacker usually distributes Trojan Horses that contain malicious code that allows the attacker to control their system. Such malicious  code is also referred to as a Backdoor. Once these Trojan Horses are executed, they may use email to inform the attacker that the system can be remotely controlled. The attacker will then install the tools required to perform the attack. Once the attacker controls enough systems, which are referred to as zombies or slaves, he or she can launch the attack[8].

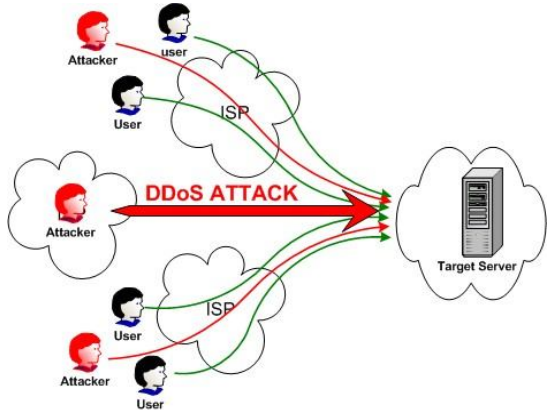

Fig1. DDoS Attack

### III.SECURE SERVICE OVERLAY (SOS) Model

The goal of the SOS architecture is to allow communication between a *confirmed user* and a *target*. The model proposes a proactive approach to prevent DoS attacks. A target is protected by removing all incoming packets from unapproved sources.   A network of selected nodes form an overlay which protect a specific target. Packets are validated at entry points of the overlay and once inside are tunneled securely to secretly designated nodes. Once validated, all traffic is forwarded to the target through the overlay[1].

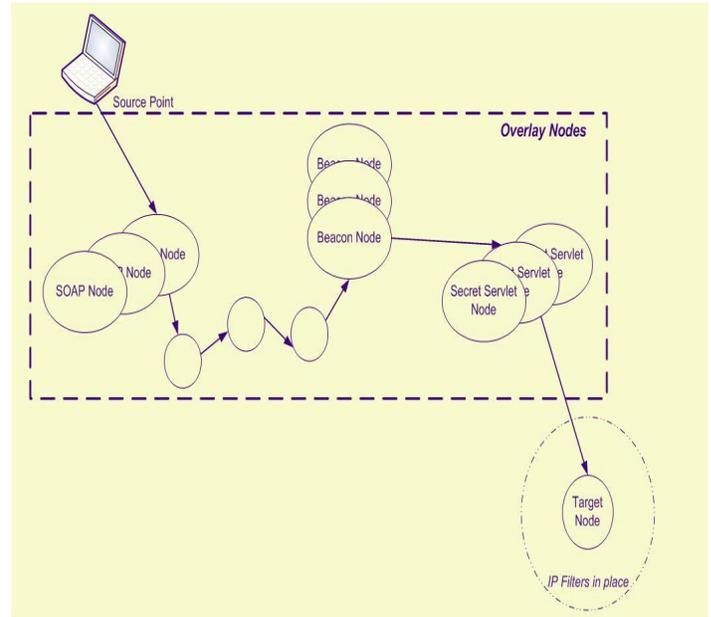

Fig2. SOS Architecture

### A.SOAP - Secure Overlay Access Point :

○   The start point in for all traffic that will communicate with the target.

○   Handles authentication of users and traffic.

### B.Target :

○   The nodes or set of nodes that will be filtered to only allow overlay traffic.

### C.Beacon :

○   The end-point in a chord ring.

○   Beacon forwards traffic to the Secret Servlet.

### D.Secret Servlet :

○   The node that will communicate with a specific target or group of targets.

### E.To mitigate attacks :

○   No unauthenticated traffic is allowed in the overlay.

○   Filtering of non-overlay traffic near the target can be done at line-speed.

○   The vulnerability of the target is offloaded onto the overlay.

○   The overlay is recoverable

### F. Design Rationale :

Fundamentally, the goal of the SOS infrastructure is to distinguish between authorized and unauthorized traffic. The former is allowed to reach the destination, while the latter is dropped or is rate-limited. Thus, at a very basic level, we need the functionality of a firewall within the network so that the access link to the  target is not congested. This imaginary firewall would perform access control by using protocols such as IPsec.

## G. Architecture Overview :

The forwarding of a packet within the SOS architecture, depicted in Figure above, proceeds through following stages:

- A source point that is the origin of the traffic forwards a packet to a special overlay node called a SOAP that receives and verifies that the source point has a legitimate communication for the target.
- The SOAP routes the packet to a special node in the SOS architecture that is easily reached, called the beacon.
- The beacon forwards the packet to a "secret" node, called the secret servlet, whose identity is known to only a small subset of participants in the SOS architecture.
- The secret servlet forwards the packet to the target.The filter around the target stops all traffic from reaching the target except for traffic that is forwarded from a point whose IP address is the secret servlet[3].

## H. Protecting the Target: Filtering :

In the current Internet, knowledge of the target's network identifier (IP address) allows an attacker to bombard the target location with packets that originate from compromised locations throughout the Internet. To prevent these attacks, a *filter* can be constructed that drops illegitimate packets at some point in the network. We assume that the filter can be constructed so that attackers do not have access to routers inside the filtered region. Essentially, we assume that the filter can be constructed locally around the target to prevent a bombardment of illegitimate traffic, while at the same time allowing legitimate, filtered traffic to successfully reach the target. Such filters need to be established at the ISP's Point of Presence (POP) routers that attach to the ISP backbone.

## I. Reaching Well-filtered Target :

Under the filtering mechanism, legitimate users can reach the target by setting the filter around the target to permit only those IP addresses that contain legitimate users. This straightforward approach has two major shortcomings. First, whenever a legitimate user moves, changes IP address, or ceases to be legitimate, the filter surrounding the target must be modified. Second, the filter does not protect the target from traffic sent by an illegitimate user that resides at the same address as a legitimate user, or from an illegitimate user that has knowledge about the location of a legitimate user and spoofs the source address of its own transmissions to be that of the legitimate user. A first step in our solution is to have the target select a subset of nodes, Ns, that participate in the SOS overlay to act as *forwarding proxies*. The filter only allows packets whose source address matches the address of some overlay node n ∈ Ns. Since n is a willing overlay participant, it is allowed to perform more complex verification procedures than simple address filtering and use more sophisticated techniques to verify whether or not a packet sent to it originated from a legitimate user of a particular target. The filtering function that is applied to a packet or flow can have various levels of complexity. It is, however, sufficient to filter on the source address: the router only needs to let through packets from one of the few forwarding proxies. All other traffic can be dropped, or rate-limited. Because of the small number of such filter rules and their simple nature (source IP address filtering), router performance will not be impaired, even if we do not utilize specialized hardware. This architecture prevents attackers with knowledge of legitimate users' IP addresses from attacking the target. However, an attacker with knowledge of the IP address of the proxy can still launch two forms of attacks: an attacker can breach the filter and attack the target by spoofing the source address of the proxy, or attack the proxy itself. This would prevent legitimate traffic from even reaching the proxy, cutting off communication through the overlay to the target. Our solution to this form of attack is to hide the identities of the proxies. If attackers do not know the identity of a proxy, they cannot mount either form of attack mentioned above unless they successfully guess a proxy's identity. We refer to these "hidden" proxies as secret servlets[6].

## J. Reaching a Secret Servlet :

To activate a secret servlet, the target sends a message to the overlay node that it chooses to be a secret servlet, informing that node of its task. Hence, if a packet reaches a secret servlet and is subsequently verified as coming from a legitimate user, the secret servlet can then forward the packet through the filter to the target.

## K. Connecting to the Overlay :

Legitimate users need not reside at nodes that participate in SOS. Hence, SOS must support a mechanism that allows legitimate traffic to access the overlay. For this purpose, we define a secure overlay access point (SOAP). A SOAP is a node that will receive packets that have not yet been verified as legitimate, and perform this verification. Effectively, SOS becomes a large distributed firewall that discriminates between authorized traffic from unauthorized traffic. Having a large number of SOAPs increases the robustness of the architecture to attacks, but complicates the job of distributing the security information that is used to determine the legitimacy of a transmission toward the target.

## L. Routing through the Overlay :

Having each overlay participant select the next node at random is sufficient to eventually reach a secret servlet. However, it is rather inefficient, with the expected number of intermediate overlay nodes contacted being $O(N=Ns)$ where N is the number of nodes in the overlay and Ns is the number of secret servlets for a particular target. Here, we discuss an alternative routing strategy in which, with only one additional node knowing the identity of the secret servlet, the route from a SOAP to the secret servlet has an expected path length that is $O(logN)$. We use Chord, which can be viewed as a routing service that can be implemented atop the existing IP network fabric, *i.e.*, as a network overlay. Consistent hashing is used to map an arbitrary identifier to a unique destination node that is an active member of the overlay.

### M. Connecting to the Overlay :

Legitimate users need not reside at nodes that participate in SOS. Hence, SOS must support a mechanism that allows legitimate traffic to access the overlay. For this purpose, we define a secure overlay access point (SOAP). A SOAP is a node that will receive packets that have not yet been verified as legitimate, and perform this verification.[2] Allowing a large number of overlay nodes to act as SOAPs increases the bandwidth resources that an attacker must obtain to prevent legitimate traffic from accessing the overlay. Effectively, SOS becomes a large distributed firewall that discriminates between authorized traffic from unauthorized traffic.

### IV.SECURITY ANALYSIS

In this section we develop simple analytical models to evaluate the performance of SOS considering DoS attacks. We make certain assumptions: an attacker knows the set of nodes that form the overlay, and can attack these nodes by bombarding them with traffic. However, the attacker does not know the precise functionality of the nodes, nor can it infer them. The bandwidth available to the attacker to launch attack upon the overlay and the target has an upper bound. Furthermore, we assume that the attackers have not breached the security protocols of the overlay, *i.e.,* their packets can always be identified by SOS as being illegitimate. Finally, each legitimate user can access the overlay through a limited number of SOAPs, but different users access  the overlay through different SOAPs.

Following are some security threats to SOS model :

### A. Denial of Service (DoS) Attacks :

*DoS attacks are increasingly mounted by professional attackers using huge zombie nets consisting of thousands of compromised machines on the Internet. Countering DoS attacks on online services has become a very challenging problem. The overlay network service provider has to protect the applications data hosted by the overlay nodes from DoS and host compromise attacks. Protecting the overlay network nodes from DoS and host compromise attacks improves service availability.*

### B. Authenticity Attacks :

*The overlay network service provider has to protect the applications data hosted by the overlay nodes from incorrect or fake (spoofed) application data. Protecting the overlay network nodes from in- correct or fake application data guarantees the authenticity of application data hosted by the nodes.*

### C. Confidentiality and Integrity Attacks :

*The overlay network service model has to protect the confidentiality and integrity from: (a) the overlay network nodes, and (b) unauthorized users.*

### V. IMPLEMENTATION OF SOS MODEL

One particularly attractive feature of the SOS architecture is that it can be implemented using existing software and standardized protocols, making its adoption and eventual use easier.

### A.Filtering:

All high and medium-range (both in terms of performance and price) routers, as well as most desktop and server operating systems, offer some high-speed packet classification scheme that can be used to implement the target perimeter filtering. A simplified version of can be used by the target to inform its perimeter routers of changes in the set of allowed secret servlets.

### B.Tunneling:

Once traffic has entered the overlay network, it needs to be forwarded to other SOS nodes toward the beacon, and from there to the secret servlets. Standard traffic tunneling techniques,and protocols can be used to this end: IP-in-IP encapsulation, GRE encapsulation, or IPsec in "tunnel mode". Furthermore, traffic inside the overlay network can take advantage of traffic prioritization schemes such as MPLS or DiffServ, if they are made available by the infrastructure providers. The routing decisions inside the overlay network are based on a Chord-like mechanism. We envision the overlay nodes to be a mix of routers and highspeed end systems. In particular, since IP tunneling is a lightweight operation, it is conceivable that SOS functionality can be offered by service providers without adversely affecting the performance of their networks. The access points to the access network can be a mix of routers and high-speed end systems. The access points and secret servlets can also act as "charging" points, if SOSlike functionality is offered on a commercial basis. Finally, since overlay nodes are only called upon to do encapsulated-packet for forwarding, cross-provider collaboration6 is a straightforward proposition, compared to controlled exposure of the filtering mechanism among different providers.

### VI.PERFORMANCE ANALYSIS OF SOS MODEL

Here we describe DoS attacks, and evaluate the SOS architecture using this model to mitigate DoS attacks. Our evaluation makes the following assumptions:

- O An attacker knows the set of nodes that form the overlay, and can attack these nodes by flooding them with traffic[3].

- O An attacker does not know which nodes are secret servlets or beacons, and does not infer these identities.

- O Attackers have not breached the security protocols of the overlay, *i.e.,* their packets can always be identified by SOS nodes as being invalid.

Each valid user can access the overlay through a limited number of SOAPs, but different users access the overlay through different SOAPs[1]. Thus, an attacker that wants to prevent all communication to the target will not target the SOAPs of a specific user, since doing so only ensures that only that user cannot communicate with the target.

## VII.CONCLUSION

Our contributions in this paper are (a) systematically studying the existing SOS architecture from the perspective of its basic design features, (b) proposing a generalized SOS architecture by introducing flexibility to the design features, (c) defining two intelligent DDoS attack models and developing an analytical approach towards analyzing the generalized SOS architecture under these two attack modelsWe addressed the problem of securing a communication service on top of the existing IP infrastructure from DoS attacks. In order to decrease the probability of a successful attack, it is desirable to have no relationship between the nodes serving special purposes in the overlay. We attack the problem with a proactive mechanism, an overlay network that can self-heal during (and after) a DoS attack, and a scalable access control mechanism that allows legitimate users to use the overlay network. We call this architecture *Secure Overlay Services*, or *SOS*. The resistance of a SOS network against DoS attacks increases greatly with the number of nodes that participate in the overlay. We believe that our approach is a novel and powerful way of countering DoS attacks, especially in service-critical environments. The SOS we have developed provide a range of defenses that can severely limit the damage caused by DoS and DDoS attacks. This is a significant step forward in providing a robust Internet service that can be used with confidence for electronic commerce and other on-line services.

## ACKNOWLEDGEMENT

This paper has benefited from conversations with many different people – far more than can be acknowledged completely here. Still we would like to particularly thank , Computer Technology Department , for guidance and support.